\documentstyle{mn}
\input{epsf.sty}
\def\pageoffset#1#2{\hoffset=#1\relax\voffset=#2\relax} 
\pageoffset{0pc}{3.5pc} 
\def\Mpc{{\,h^{-1}\,{\rm Mpc}}}

\def\and  {\it {et al.} \rm}

\def\etal{{\rm et~al. }}
\def\spose#1{\hbox to 0pt{#1\hss}}
\def\simlt{\mathrel{\spose{\lower 3pt\hbox{$\mathchar"218$}}
     \raise 2.0pt\hbox{$\mathchar"13C$}}}
\def\simgt{\mathrel{\spose{\lower 3pt\hbox{$\mathchar"218$}}
     \raise 2.0pt\hbox{$\mathchar"13E$}}}
\def\beq{\begin{equation}}
\def\eeq{\end{equation}}
\def\bce{\begin{center}}
\def\ece{\end{center}}
\def\bea{\begin{eqnarray}}
\def\eea{\end{eqnarray}}
\def\ben{\begin{enumerate}}
\def\een{\end{enumerate}}

\def\ni{\noindent}
\def\nn{\nonumber}

\def\brr{\begin{array}}
\def\err{\end{array}}

\def\etal{{\rm et~al. }}

\def\nh1{n_{\rm HI}}

\def\k{{\cal K}}
\def\N{{\cal N}}

\def\p1dk{P_{\rm 1D}(k)}
\def\simlt{\mathrel{\spose{\lower 3pt\hbox{$\mathchar"218$}}
     \raise 2.0pt\hbox{$\mathchar"13C$}}}
\def\simgt{\mathrel{\spose{\lower 3pt\hbox{$\mathchar"218$}}
     \raise 2.0pt\hbox{$\mathchar"13E$}}}

\newcommand{\Nbar}{\overline{N}}

\newcommand{\xibar}{\overline{\xi}}

\def\Or{{\cal O}}

\def\calG{{\cal G}}

\newcommand{\de}{\delta}
\newcommand{\lexp}{\mathop{\bigl\langle}}
\newcommand{\rexp}{\mathop{\bigr\rangle}}

\begin{document}
\title[2-Point Moments I: Theory and Simulations]
{2-Point Moments in Cosmological Large Scale Structure  
\\ I.  Theory and Comparison with Simulations}

%%%%%%%%%%%%%%%%%%%%%%%%%%%%%%%%%%%%%%%%%%%%%%%%%%%%%%%%%%%%%%%%%%%%%%%%%%%%%%%
%%%%%%%%%%             AUTHORS       %%%%%%%%%%%%%%%%%%%%%%%%%%%%%%%%%%%%%%%%%%
\author[]
%[\ni E. Gazta\~{n}aga, P.Fosalba, R.A.C. Croft]
{\ni E. Gazta\~{n}aga$^{1,2}$,  P.Fosalba$^{3,4}$,  R.A.C. Croft$^{5}$ \\
$^{1}$ Instituto Nacional de Astrof\'{\i}sica, \'Optica y Electr\'onica
(INAOE), Tonantzintla, Apdo. Postal 216 y 51, 7200 Puebla, Mexico \\
$^{2}$ Institut d'Estudis Espacials de Catalunya  (IEEC/CSIC), 
Edifici Nexus-201 - c/ Gran Capit\`a 2-4, 08034 Barcelona, Spain \\
$^{3}$ Astrophysics, Space Science Department 
of ESA, ESTEC, NL-2200 AG Noordwijk, The Netherlands \\
$^{4}$ Institut d'Astrophysique de Paris,
98bis, boulevard Arago, F-75014 Paris, France \\
$^{5}$ Harvard-Smithsonian Center for Astrophysics, 60 Garden St,
Cambridge, MA 02138, USA}

\maketitle

%%%%%%%%%%%%%%%%%%%%%%%%%%%%%%%%%%%%%%%%%%%%%%%%%%%%%%%%%%%%%%%%%%%%%%%%%%%%%%%

\begin{abstract}
  
  We present new perturbation theory (PT) predictions in the Spherical
  Collapse (SC) model for the 2-point moments of the large-scale distribution
  of dark matter density in the universe.  We assume that these fluctuations
  grow under gravity from small Gaussian initial conditions.  These
  predictions are compared with numerical simulations and with previous PT
  results to assess their domain of validity.  We find that the SC model
  provides in practice a more accurate description of 2-point moments than
  previous tree-level PT calculations. The agreement with simulations is
  excellent for a wide range of scales ($5-50 \Mpc$) and fluctuations
  amplitudes ($\sigma^2\simeq 0.02-2$).  When normalized to unit variance
  these results are independent of the cosmological parameters and of the
  initial amplitude of fluctuations.  The 2-point moments provide a convenient
  tool to study the statistical properties of gravitational clustering for
  fairly non-linear scales and complicated survey geometries, such as those
  probing the clustering of the Ly$\alpha$ forest.  In this context, the
  perturbative SC predictions presented here, provide a simple and novel way
  to test the gravitational instability paradigm.

\end{abstract}
%%%%%%%%%%%%%%%%%%%%%%%%%%%%%%%%%%%%%%%%%%%%%%%%%%%%%%%%%%%%%%%%%%%%%%%%%%%%%%%

\section{Introduction}
\label{sec:intro}

The 2-point point correlation function $\xi_2$ is a well known and useful (eg
Peebles 1980, 1993) characterization of fluctuations in the large scale
density distribution. It measures the covariance of density fluctuations
$\xi_2(r_1,r_2) \equiv <\delta(r_1)\delta(r_2)>$ and should be translationally
invariant and isotropic, i.e, $\xi_2(r_1,r_2)=\xi_2(r_{12})\equiv|r_2-r_1|)$,
in an homogeneous and isotropic distribution. In fact, for a Gaussian field,
$\xi_2(r_{12})$ is all we need to characterize fluctuations, as all higher
order correlations $\xi_N(r_1,\dots,r_N) \equiv
<\delta(r_1)\dots\delta(r_N)>_c$ are zero (the subscript $c$ stands for the
'connected' or reduced moments). Even if the initial conditions were Gaussian,
non-linear evolution of the density field under gravity introduces
non-Gaussian correlations so that both in the weakly non-linear regime (e.g,
Fry 1984) or the strongly non-linear regime (e.g, Davis \& Peebles 1977) we
expect $\xi_N \simeq \xi_2^{N-1}$.  It is therefore important to characterize
and predict in detail the behavior of higher order correlations.  One option
is to consider $\xi_N$ itself, but this is complicated because of the
multimensional space $r_1,\dots,r_N$ involved.  Another option is to consider
N-order one-point moments or cumulants $\k_N \equiv <\delta^N>$, smoothed over
a fixed scale $R$.  This has the advantage of providing good signal to noise
in the measurements (as we are averaging over multidimensional space) but at
the cost of losing configuration information. Another problem is to take into
account the smoothing window when we have complicated geometries. For example
when the survey consists of one dimensional pencil beams, as is the case of
the density traced by the Ly$\alpha$ forest in QSO spectra.  This is in fact
the main physical applications that motivated the present study.

Intermediate between cumulants and N-point correlations are
2-point cumulants: 
\beq
\k_{p q}(r_1,r_2) \equiv <\delta(r_1)^p\delta(r_2)^q>_c
\eeq
where $\k_{p q}(r_1,r_2)=\k_{p q}(r_{12})$, in a homogeneous and isotropic
distribution. These 2-point cumulants have some of the advantages and
disadvantages of N-point functions and N-order cumulants.  They do contain
information on higher-order correlations ($\k_{p q}=0$, for $p+q>2$ in a
Gaussian field), but are better suited than cumulants for situations
where we want to avoid boundary restrictions in complicated geometries.
 As mentioned previously, a good example is
the Ly$\alpha$ forest, 
which traces a narrow one dimensional sample of the density
distribution along the line-of-sight of the QSO.

As in the case of the cumulants and N-point functions one can use perturbation
theory (PT) to predict the values of $\k_{p q}$ in gravitational perturbation
theory from some given (typically Gaussian) initial conditions. 
Predictions for the 2-point moments in the PT approach to lowest
order were first
derived by Bernardeau (1996; B96 hereafter).  

This paper is motivated in part by the lack
of measurements of 2-point moments,  $\k_{p q}$, from numerical simulations. 
As pointed out in B96, numerical results for  $\k_{p q}$
are difficult to obtain in a regime where they can be directly 
compared with the PT predictions. However, simulation work is an
important and necessary step, as we need to study both 
the non-linear regime and the range of
validity of the PT calculations for the 2-point statistics. 

As we will show below this comparison 
can be most conveniently done by computing the PT predictions for the
density moments using the Spherical Collapse  (SC) model.
The SC model has been shown to provide 
a simple and accurate way of predicting 
the 1-point statistics of gravitational clustering both in the weakly
non-linear (Bernardeau 1992, 1994; Fosalba \& Gazta\~naga 1998,
FG98 hereafter; Gazta\~naga \& Fosalba 1998)
and fully non-linear  
(see e.g, Gazta\~naga \& Croft 1999; Gazta\~naga \& Scherrer 2001) 
regimes.

In this paper we concentrate on the perturbative approach.  The fully
non-linear case and applications to  Ly$\alpha$  forest will be presented
elsewhere (Gazta\~naga \etal 2001).  In paper II of this series we present
results in redshift space and a comparison with clustering in galaxy data.

This paper is organized as follows. In \S\ref{sec:2pdf} we introduce the
2-point statistics. Perturbative predictions in the SC model are given in
\S\ref{sec:ptresults} and a detailed comparison to Nbody simulations is
presented in \S\ref{sec:nbody}.  Finally, we summarize our results and give
our conclusions in \S\ref{sec:conc}.

\section{The 2-point PDF and its Cumulants}
\label{sec:2pdf}

\subsection{The PDF of the Initial Conditions}

In the limit of early times, we assume a nearly homogeneous matter
distribution with very small fluctuations (or seeds), with given statistical
properties.  We will concentrate on the case where the statistics of the
initial density are well described by Gaussian initial conditions, which
correspond to a broad class of physical models to generate initial conditions.

The 2-point PDF  is given by a bivariate normal
distribution for  $P_G(\de_1,\de_2)$:

\beq
P_{12}^G(\de_1,\de_2) ~ = ~{1\over{2\pi \sqrt{det C}}}~ \exp{\left[-{1\over{2}} \sum_{i,j} 
~\de_i ~c_{ij}^{-1} ~\de_j \right]},
\eeq
where $C$ is the inverse of the covariance matrix: 
\beq
c_{ij} \equiv ~ <\delta_i~\delta_j>  ~~~~~~ i,j= 1 2.
\eeq
In our case we have that $c_{1 1}=c_{2 2} \equiv \sigma^2$,
and  $c_{12}=c_{21} \equiv \xi_2(r_{12})$ is
the 2-point correlation function. Notice that $\sigma^2=\xi_2(0)$.
We then have:
\beq
P_{12}^G~ = ~{1\over{2\pi \sqrt{\sigma^4-\xi_2^2}}}~ \exp{\left[-{\sigma^2 \de_1^2+\sigma^2 \de_2^2
-2 \xi_2 \de_1 \de_2 \over{2(\sigma^4-\xi_2^2)}}\right]}
\label{p12g}
\eeq
As correlations
decrease with separation,
we typically have $\xi_2(r_{12}) << \sigma^2$. In this limit
we find:
\beq
P_{12}^G(\de_1,\de_2)~\approx ~P_1^G(\de_1)~ P_1^G(\de_2) ~
 \left[ 1 + {\xi_2(r_{12})\over{\sigma^2}} ~ 
{\de_1\de_2\over{\sigma^2}}\right],
\label{p12gic}
\eeq
where $P_1^G(\de)$ is the 1-point PDF for a Gaussian field
\footnote{Note that a Gaussian PDF
only makes physical sense, i.e, $\delta>-1$ ($\rho>0$),
when the variance is small: $\sigma \rightarrow 0$}.
Thus, the explansion parameter is $\alpha \equiv
\xi_2(r_{12}/\sigma^4$, and we have:
\beq
P_{12}^G(\de_1,\de_2)~\approx ~P_1^G(\de_1)~ P_1^G(\de_2) ~
 \left[ 1 +\alpha ~ \de_1\de_2 +  \Or\left(\alpha^2\right) \right].
\label{p12gic2}
\eeq
As we shall see below (see \S\ref{sec:ptresults})
the expansion in Eq.(\ref{p12gic2}) is a basic ingredient for
generating PT predictions  
for the 2-points moments, as it simply factorizes the 2-point
average into products of 1-point averages.

\subsection{The Evolved Mass PDF}
\label{masspdf}

\subsubsection{Linear Evolution}

Because of gravitational growth, the evolution of the matter density 
field, $\delta$, will change the initial PDF.  
For small fluctuations linear theory provides a
simple prediction for the time evolution: 
\beq \delta(t,x)= D(t)
~\delta_0(x) \equiv \delta_L, 
\eeq 
where $D(t)$ is the growth factor
(equal to the scale factor $D=a$ for $\Omega=1$), and $\delta_0(x)$ is
the initial field. We will denote this linear prediction by
$\delta_L$. For Gaussian initial conditions the linear PDF is also
Gaussian with a 2-point function $\xi_2(t)$, given by scaling the
initial value $\xi_2(t_0)$ by $D^2(t)$, so that:
\beq
\xi_2(t)= D^2(t)~ \xi_2(t_0).
\eeq

\subsubsection{Non-linear Evolution}

The (Newtonian) non-linear equations of motion of density fluctuations in the 
matter dominated regime have no exact analytic solution (Peebles 1980). 
However, one can conveniently
describe the non-linear evolution of density fluctuations by making use
of the Spherical Collapse approximation (SC hereafter).
Within this approximation
the non-linear overdensity $\rho \equiv 1+\delta$\footnote{As it is  
more convenient, in the equations below we will use the notation 
 $1+\delta$ for the non-linear case and $\delta_L$
for the linear one.}  in 
{\em Lagrangian} space is simply related to the linear one by:
\beq
\rho(q) = {\calG} [\delta_L(q)] 
\label{eq:deltasc}
\eeq
where $\delta_L(q) \equiv D \delta_0(q)$, and we stress that 
we choose Lagrangian coordinates $q$ to describe the growth of 
fluctuations. The non-linear 2-point probability  
distribution function of matter density fluctuations, 
$P(\de_1,\de_2)$, is thus related to the  
linear (Gaussian) one 
by a simple change of variable: 
\beq 
P(\rho_1,\rho_2) = P_G(\de_1^L,\de_2^L) ~ 
\left| {d\de_1^L\over{d\rho_1}} \right| 
\left| {d\de_2^L\over{d\rho_2}} \right| 
\label{rhopdf} 
\eeq 
where $\de_L=\calG^{-1}[\rho]$.  
 
As we are interested in the {\em weakly non-linear} 
regime, we shall concentrate on the perturbative (or Taylor) expansion 
of the non-linear density fluctuation in terms of the linear one:
\beq
\rho(q) = \calG[\delta_L(q)] = \sum_n {\nu_n\over{n!}} ~[\delta_L(q)]^n
\label{eq:mapping}
\eeq
where the $\nu$ coefficients specify the non-linear coupling between
density fluctuations in real-space. Note that in the SC, these $\nu$'s
are the equivalent to the angle average (spherically symmetric
component) of the couplings between modes as given by the PT kernels 
in Fourier space (see FG98 for details).
The first such coefficients are (see Bernardeau 1992, 1994; FG98),
\bea 
\nu_2 &=& {34\over 21} \sim 1.62 \nn \\  
\nu_3 &=& {682\over 189} \sim 3.61 \nn \\  
\nu_4 &=& {446440\over 43659} \sim 10.22 \nn \\ 
\nu_5 &=& {8546480\over 243243} \sim 35.13  
\label{nusc} 
\eea

\subsubsection{Lagrangian to Eulerian PDF} 
\label{LtoEpdf} 
 
As mentioned before,  
the above expression Eq.(\ref{rhopdf}) 
for the PDF corresponds to the probability distribution 
of the evolved field in Lagrangian space, $q$. This corresponds
to a fixed mass, while we want the statistics for a fixed volume. 
The Lagrangian and Eulerian elements can be related 
by imposing mass conservation: 
\beq 
d\de_1(x) ~d\de_2(x) =  
\rho_1~\rho_2~ d\de_1(q)~ d\de_2(q) , 
\eeq 
where $\rho(x)=1+\de(x)$ is the overdensity in Eulerian coordinates. 
We then have: 
\beq 
P_E(\rho_1,\rho_2) = {1\over{\rho_1\rho_2}}~P_L(\rho_1,\rho_2) 
\eeq 
up to a normalization constant. 
The Eulerian PDF is therefore 
\beq 
P_E(\rho_1,\rho_2) = {1\over{N}} {P_G(\de_1^L,\de_2^L)\over{\rho_1\rho_2}} ~ 
\left| {d\de_1^L\over{d\rho_1}} \right| 
\left| {d\de_2^L\over{d\rho_2}} \right|, 
\label{p12nla} 
\eeq 
where $N$ is a normalization constant.  
 
In the perturbative regime one finds that the PDF in Euler and
Lagrange space are the same, as the term $\rho_1\rho_2$ in 
the denominator above only introduces changes in higher order
corrections (i.e, to leading order, $\rho_1\rho_2 \simeq 1$). 
The next-to-leading order in the expansion of the PDF or its 
moments, the so-called {\em one-loop}
correction (see Scoccimarro \& Frieman 1996a,b) has a contribution 
from the  mass conservation factor, $\rho_1\rho_2 \simeq \rho_1$, 
which yields,
\beq 
P_E(\rho_1,\rho_2) \simeq {1\over{N}} {P_G(\de_1^L,\de_2^L)\over{\rho_1}} ~ 
\left| {d\de_1^L\over{d\rho_1}} \right| 
\left| {d\de_2^L\over{d\rho_2}} \right|, 
\label{p12nl} 
\eeq 
and which we shall use below for the estimation of the 2-point moments of the
dark matter density field in the weakly non-linear regime
\footnote{
This expression apparently breaks the $1 \rightarrow 2$ symmetry, but 
this is only a trick that will be useful for simplifying
 the moment estimations  
below (see also the argument in \S3.2.2 of B96)}.

\subsection{2-point Cumulants} 
 
Consider now a generic field, $\phi$, which could be either 
the measured flux in a 1D quasar spectrum or the mass density in 3D  
space. We will assume that this field has been smoothed over 
some resolution cell $\lambda$. We define 
the reduced 2-point reduced moments or 
cumulants $\k_{p q}$ of the field $\phi$ 
at two different positions 
 $r_1$ and $r_2$, by: 
 
\beq 
\k_{p q} \equiv <\phi(r_1)^p \phi(r_2)^q >_c = \left. {{\partial^2{ \log M[t_1,t_2]}} 
\over{\partial t_1 \partial t_2}}\right|_{t_1,t_2 \rightarrow 0}, 
\label{kijdef} 
\eeq 
where $M[t_1,t_2]=<\exp(\phi_1t_1)\exp(\phi_2t_2)>$  
is the joint generating function of the  
(un-reduced) moments: 
\beq 
m_{p q} \equiv  <\phi(r_1)^p \phi(r_2)^q > = 
 \left. {{\partial^2{{M[t_1,t_2]}}} 
\over{\partial t_1 \partial t_2}}\right|_{t_1,t_2 \rightarrow 0}, 
\eeq 
which can also be readily obtained by using the corresponding 
2-point probability distribution: 
 
\beq 
m_{p q} =  <\phi(r_1)^p \phi(r_2)^q > =
\int P(\phi_1,\phi_2) ~  \phi_1^p \phi_2^q ~ d\phi_1 d\phi_2  
\eeq 
 
For example, the 2-point function of the field 
is just given by: 
 
\beq 
\xi_2(r_{12}) \equiv \k_{1 1}(r_{12}) = <\phi_1 \phi_2>_c. 
\eeq 
Of course, as mentioned in the introduction, these 2-point cumulants are 
just a function of the separation $r_{12} = r_2-r_1$ 
between the cells: 
$\k_{p q}= \k_{p q}(r_{12})$ .
 
The first reduced moments are: 
\bea 
\k_{0 0} &=& m_{0 0} \\ 
\k_{0 1} &=& m_{0 1}  \nn \\ 
\k_{0 2} &=& \sigma^2 = m_{0 2} - m_{0 1}^2 \nn \\ 
\k_{0 3} &=& m_{0 3} -3 m_{0 1} m_{0 2}+ 2 m_{0 1}^3  \nn \\ 
\k_{0 4} &=& m_{0 4}- 6 m_{0 1}^4 +12 m_{0 1}^2m_{2 0}  
     - 3 m_{2 0}^2 -4 m_{0 1}m_{0 3} \nn \\ 
\k_{1 1} &=& \xi_2 = m_{1 1} - m_{0 1}^2 \nn \\ 
\k_{1 2} &=& m_{1 2}- m_{0 1} m_{0 2}-2 m_{0 1} m_{1 1} + 2 m_{0 1}^3 \nn  
%%\k_{2 2} &=& m_{2 2} -2  m_{1 1}^2 - m_{2 0}  m_{0 2} \nn  
%%\k_{1 3} &=& m_{1 3} -3 m_{0 2} m_{1 1} \nn \\ 
%%\k_{1 4} &=& m_{1 4} -4 m_{0 3} m_{1 1} -6 m_{0 2} m_{1 2}\nn \\ 
%%\k_{2 3} &=& m_{2 3} -6 m_{1 2} m_{1 1} -3 m_{0 2} m_{2 1} - m_{0 3} m_{2 0} \nn 
\label{kijmij} 
\eea 
and so on. Note that even when we normalise the field
 so that the 
mean is zero ($m_{0 1}=0$), the cumulants are 
different from the central moments in  
that they have the lower order 
moments subtracted, i.e: 
\bea
\k_{2 2} &=& m_{2 2} -2  m_{1 1}^2 - m_{2 0}  m_{0 2} \nn \\
\k_{1 3} &=& m_{1 3} -3 m_{0 2} m_{1 1} \nn \\ 
\k_{1 4} &=& m_{1 4} -4 m_{0 3} m_{1 1} -6 m_{0 2} m_{1 2}\nn \\ 
\k_{2 3} &=& m_{2 3} -6 m_{1 2} m_{1 1} -3 m_{0 2} m_{2 1} - m_{0 3}m_{2 0} 
\label{kij}
\eea
for $m_{0 1}=0$.  
 
It is interesting to define the following 1-point hierarchical 
ratios: 
\beq 
S_q = {\k_{0 q}\over{\k_{0 2}^{q-1}}} ~~~~~~~~ q>2 
\label{sj} 
\eeq 
and the corresponding 2-point generalization: 
\beq 
c_{p q} = {\k_{p q}\over{\k_{1 1}~ \k_{0 2}^{p+q-2}}}
 = {\k_{p q}(r_{12})\over{~\xi_2(r_{12}) ~\sigma^{2(p+q-2)}}}  \quad , \quad
p+q>2 
\label{qij} 
\eeq 
At $r_{12} = 0$ the 2-point cumulants become 1-point cumulants,
$\k_{i j} \rightarrow \k_{i+j}$
and consequently, for the reduced moments, $c_{p q} \rightarrow S_{p+q}$. 
These ratios  turn out to be roughly constant under 
gravitational evolution from Gaussian initial conditions (see Peebles 1980).

\subsection{Discreteness Effects}
\label{sec:disc}

For discrete fields, such as fluctuations traced by 
the galaxy distribution or particles in Nbody simulations, we  must
correct the cumulants to account for Poisson fluctuations around
the mean density $\Nbar \equiv <N>$ at the smoothing scale. 
Following Gazta\~naga \& Yokoyama (1993), there is a simple relation between
the discrete cumulant generating function $\N^D(t)$ and the
corresponding continuous one $\N^C(t)$:
\beq
\N^D [t] = \N^C [e^t-1],
\eeq
where $\N(t)$ generates the cumulants (or reduced moments) of the
density field: $N(r)$. The latter is simply related to the density 
fluctuation $\delta(r)$, $N(r) = \Nbar [1+\delta(r)] $. 
We can trivially extend the above
relation to the 2-point case:
\beq
\N^D [t_1,t_2] = \N^C \left[e^t_1-1, e^t_2-1\right].
\eeq
This relation yields the following corrections for estimating the 
continuous cumulants $\k_{p q}^C$ from the measured, discrete, ones
$\k_{p q}^D$:
\bea 
\k_{0 2}^C &=& \k_{0 2}^D - {1\over{\Nbar}}, \\ 
\k_{0 3}^C &=& \k_{0 3}^D - {3\over{\Nbar}} \k_{0 2}^C - {1\over{\Nbar^2}}\nn, \\ 
\k_{1 1}^C &=& \k_{1 1}^D \nn \\ 
\k_{1 2}^C &=& \k_{1 2}^D - {1\over{\Nbar}} \k_{1 1}^C  \nn, \\ 
\k_{1 3}^C &=& \k_{1 3}^D - {3\over{\Nbar}} \k_{1 2}^C - {1\over{\Nbar^2}} \k_{1 1}^C  \nn, \\ 
\k_{2 2}^C &=& \k_{2 2}^D - {2\over{\Nbar}} \k_{1 2}^C -  {1\over{\Nbar^2}} \k_{1 1}^C
 \nn ,
\eea
and so on. In all cases $\k_{p q}=\k_{p q}(r_1,r_2)$ with $r_1\ne r_2$.
Note how the two point function $\k_{1 1}$ is not affected by
Poisson fluctuations, unless the two points are the same $r_1= r_2$
(e.g, $\k_{0 2}$). This reflects the fact that Poisson fluctuations 
are not spatially 
correlated and it only yield contributions when we have two or more cells
at the same location.

In the limit when shot-noise dominates the cumulants we have
that the discrete fields give for Eq.(\ref{sj}):
\beq 
S_q^{Poisson} =  {\k_{0,q}^D\over{(\k_{0 2}^D)^{q-1}}} = 1 \quad ,
\quad  q>2 .
\label{sjp} 
\eeq 
In this same limit $c_{p q}^{Poisson}=0$. For a Gaussian continuous field
(i.e, neglecting only the clustering of higher orders $p+q>2$ ), we find:
\bea 
c_{1 2}^{Poisson} &=& 1 \qquad ; \nn \\ 
c_{2 2}^{Poisson} &=& 3 \quad , \quad c_{1 3}^{Poisson} = 4  ~; \nn \\ 
c_{2 3}^{Poisson} &=& 18 \quad , \quad c_{1 4}^{Poisson} = 32 ~;
\label{qijp} 
\eea
and so on. These values illustrate the fact that Poisson fluctuations 
produce artificial non-Gaussianities which mimic the
hierarchical clustering. We will see below that the hierarchical amplitudes
that emerge from gravitational growth
 (starting from Gaussian initial conditions) are 
significantly higher for all realistic situations. Nevertheless, the
shot-noise must  be subtracted (using the formulae above) if we
want an accurate comparison with predictions from gravitational instability.

\section{Perturbative Predictions} 
\label{sec:ptresults} 

\subsection{2-point Cumulants for the Mass} 
 
We can now use the 2-point quasi-linear distribution induced by the SC, i.e,
Eq.(\ref{p12nl}), to estimate $m_{p q}$ and $\k_{p q}$ for mass density
fluctuations.  We first consider the case where the resolution cell $\lambda
\rightarrow 0$.  This is a straightforward, rather tedious calculation (ideally
fitted for algebraic programming, such as Mathematica). The first step is to
expand the moments in a perturbative series. We use
Eq.(\ref{p12gic2}) and the expansion Eq.(\ref{eq:mapping}) for the non-linear 
overdensity $\calG$:

\bea 
&& m_{p q} ~\equiv~  <\calG_1^p~\calG_2^q >
            ~ =~ I_{p,0} I_{q,0} + \alpha ~ I_{p,1} I_{q,1} \nn \\
&&         ~+~
        {1\over 2} ~\alpha^2 \left[ I_{p,2} I_{q,2} -  
        ( I_{p,2} I_{q,0} + I_{q,2} I_{p,0} ) ~\sigma^2  
        + I_{p,0} I_{q,0} ~\sigma^4   
        \right]  \nn \\
&&         ~+~ 
        {1\over 6} ~\alpha^3 \left[ I_{p,3} I_{q,3} -  
        3 ~( I_{p,3} I_{q,1} + I_{q,3} I_{p,1} ) ~\sigma^2 +
\right.
\nonumber \\
&& \left.  
        9 ~ I_{p,1} I_{q,1} ~\sigma^4 \right]  ~+~
        \Or\left(\alpha^4\right)  \nonumber ~;~ \\
&& I_{m,n} ~\equiv~ \lexp (\calG-1)^m ~\delta_L^n \rexp
\label{mij} 
\eea
where we have truncated the expansion to third order in
$\alpha = {\xi_2(r_{12}) /{\sigma^4}}$,  
and the mean $<\dots>$ is taken over the 1-point Gaussian PDF 
$P_G(\delta)$. 
Note that in Lagrangian space we also need to include 
a term $1/\calG = 1/\rho$ to account for mass conservation.

\subsubsection{Results from the SC Model} 
\label{sec:lagrangian} 

As discussed in \S\ref{sec:2pdf}, 
the SC approximation is derived in {\bf Lagrange} space, but
clustering  in the simulation and observation samples is measured
in {\bf Euler} space.
To normalize appropriately the moments to Euler space, we shall 
take into account  mass conservation, as discussed in \S\ref{LtoEpdf}.    
This implies that the 2-point moments in Lagrange and Euler space 
relate to one another in the following way: 
\beq 
< \rho_1^J \rho_2^K >_E = < \rho_1^{J-1} \rho_2^K >_L ~<1/\rho >_L^{J+K-1},
\label{l2e} 
\eeq 
where the sub-indices $E$ and $L$ denote Euler and Lagrange space,  
respectively. The analogous expression for the 1-point moments is recovered 
by setting $K = 0$.  

Introducing Eq.(\ref{l2e}) into Eqs.(\ref{kij}) \& (\ref{mij}), 
one can derive PT predictions for the 2-point moments to arbitrary order.  
In particular, we find 
that all the 2-point cumulants $\k_{p q}$ (with $p,q > 0$) 
will be proportional to  
the two point function $\xi_2 \equiv \k_{1 1}$,
\beq 
\k_{p q}(r_{12}) = ~~ c_{p q}~~(\sigma^2)^{p+q-2} ~~ \xi_2(r_{12})  
+ \Or(\xi_2^2) 
\eeq 
where $\sigma^2=  \k_{1 1}(0) = \k_{0 2}=\k_{2 0}=\sigma^2(\lambda)$  
is the variance at the resolution cell. 

To leading order in the two point function, $\xi_2 \equiv \k_{1 1}$,
they are given by:
\bea 
\k_{1 1} &=& \left[1 + (3 - 3 \nu_2 + \nu_3 )\sigma^2 \right] ~\xi_2  \\ 
        &+& \left( {\nu_2^2\over 2} - \nu_2 \right)  ~\xi_2^2  + {\Or (\xi_2^3)} \nn \\ 
 \k_{0 2} &=& \sigma^2 + \left(3 - 4 \nu_2 + {1\over 2} \nu_2^2 + \nu_3 \right) ~\sigma^4 + {\Or (\sigma^6)}     
\nn \\ 
 \k_{1 2} &=& \left[ 2 \nu_2  \sigma^2 -
 \left(2 -13 \nu_2 +9 \nu_2^2 + 3 \nu_3 
- 3 \nu_2 \nu_3 - \nu_4 \right) \sigma^4 \right] ~\xi_2
\nn \\   
&+& \nu_2 ~\xi_2^2 + {\Or (\xi_2^3)} 
%\label{k12}  
\nn \\  
 \k_{0 3} &=& 3 \nu_2 \sigma^4   \nn \\
        &+& \left( -2 + 18 \nu_2 -{33\over 2} \nu_2^2 - 4\nu_3 \right. \nn \\
        &+& \left.
\nu_2^3 + 6 \nu_2 \nu_3 + {3\over 2} \nu_4 \right) \sigma^6 
        + \Or\left(\sigma^8\right) \nn   
\label{kijloopl} 
\eea 
 
Similarly, for the reduced moments (or hierarchical ratios), 
Eq.(\ref{qij}), we get,   
\bea 
 c_{1 2} &=& 2 \nu_2 + 
(-2 +\nu_2 + 5\nu_2^2 -3\nu_3   
-\nu_2^3 -\nu_2 \nu_3 + \nu_4) ~\sigma^2 \nn \\ 
&+& \left[ {\nu_2 \over \sigma^2}  
-\nu_2 +{3\over 2} \nu_2^2 + 
{1\over 2} ( -\nu_2^3 + \nu_4)  \right] ~\xi_2   
\nn \\ 
&+& \left( \nu_2^2 - \nu_3 + \nu_2 \nu_3 - {\nu_2^3 \over 2} \right) 
~{\xi_2^2\over \sigma^2}  + {\Or (\xi_2^3)} 
\label{c12scl} 
\eea
\bea
c_{1 3} &=&  6 \nu_2^2 + 3 \nu_3 \nn \\  
&+& \left( 6 - 30 \nu_2 + 18 \nu_2^2 + 30 \nu_2^3 -6 \nu_2^4 - 2 \nu_3 - 9 \nu_2 \nu_3 
\right. \nn \\ 
&-& \left. 9 \nu_2^2 \nu_3 - {3\over 2} \nu_3^2 - 4 \nu_4 +  
9 \nu_2 \nu_4 + {3\over 2} \nu_5 \right) ~\sigma^2 \nn \\ 
&+& \left[ {6 \nu_2^2 \over \sigma^2} - 6 \nu_2 - 3 \nu_2^2 + 24 \nu_2^3 - 
6 \nu_2^4 \right. \nn \\
&-& \left. 9 \nu_2 \nu_3 - {3\over 2} \nu_2^2 \nu_3 +  
{15\over 2} \nu_2 \nu_4 \right] ~\xi_2   
\nn \\ 
&+& \left[ {\nu_3 \over \sigma^2} + 6 \nu_2^3 - 3 \nu_2^4 - 2 \nu_3 -  
3 \nu_2 \nu_3 \right. \nn \\
&+& \left. 5 \nu_2^2 \nu_3 + {3\over 2} \nu_3^2 + {\nu_5\over 2} \right] 
~ {\xi_2^2\over \sigma^2} \nn \\ 
&+& \left( \nu_2 \nu_3 - {1\over 2} \nu_2^2 \nu_3 - \nu_4 +  
{3\over 2} \nu_2 \nu_4 \right)~ {\xi_2^3\over \sigma^4}   
  + {\Or (\xi_2^4)}  
\label{c13scl} 
\eea
\bea 
 c_{2 2} &=& 4 \nu_2^2 +   
\left( - 4 \nu_2 - 6 \nu_2^2 + 24 \nu_2^3 - 4 \nu_2^4 \right. \nn \\
&-& \left.  6 \nu_2 \nu_3 
-4 \nu_2^2 \nu_3 + 4 \nu_2 \nu_4 \right)~ \sigma^2 \nn \\ 
&+& \left[  
{4 \over \sigma^2} (\nu_2^2 +\nu_3) + 6 - 24 \nu_2 + 11 \nu_2^2 + 22 \nu_2^3  
\right. \nn \\ 
&-& \left. 4 \nu_2^4 - 6 \nu_3 + 6 \nu_2 \nu_3 -  
8 \nu_2^2 \nu_3 - 2 \nu_3^2  \right. \nn \\
&-& \left. 3 \nu_4 + 6 \nu_2 \nu_4 + 2 \nu_5 \right]~\xi_2 
\nn \\ 
&+& \left[ 4 {\nu_2^2\over \sigma^2} - 8 \nu_2 + 8 \nu_2^2 + 16 \nu_2^3 - 
6 \nu_2^4 - 14 \nu_2 \nu_3 \right. \nn \\
&+& \left. 2 \nu_2^2 \nu_3 - 2 \nu_4 + 8 \nu_2 \nu_4  
\right] ~ {\xi_2^2\over \sigma^2} \nn \\ 
&+& \left( 2 \nu_2^2 - 2 \nu_2^3 - \nu_2^4 + 2 \nu_3 \right. \nn \\ 
&-& \left. 6 \nu_2 \nu_3 + 
4 \nu_2^2 \nu_3 + 2 \nu_3^2 \right) ~ {\xi_2^3\over \sigma^4}  
  + {\Or (\xi_2^4)}     
\label{c22scl} 
\eea 
The above expressions, Eqs.(\ref{c12scl})-(\ref{c22scl}), are the main
analytic results of this paper.
Notice that, to leading order (tree level in PT) for 
Gaussian initial conditions, the 
following property holds (see B96): 
\beq 
c_{p q}= c_{1 p} ~c_{1 q}. 
\eeq 
However, as can be seen from the above formulae, 
Eq.(\ref{c12scl})-(\ref{c22scl}), 
this is not true beyond the leading order (i.e, for the loop
corrections in PT).  
 
In the limit $r_{12} \rightarrow 0$, there is a simple correspondence  
between the 2-point and 1-point moments $\k_{i j} \rightarrow
\k_{i+j}$, or equivalently, $c_{i j} \rightarrow S_{i+j}$.
In particular, in this limit, one recovers 
known results for the 1-point moments in the SC model 
(see FG98, Appendix A1):
\bea 
 \k_{1 1} &=& \k_{0 2} \rightarrow \sigma^2_{NL} = 
\sigma^2 + \left( 3 -4 \nu_2 + {1\over 2} \nu_2^2 + \nu_3 \right) \sigma^4  
        + \Or\left(\sigma^6\right) \nn  \\
\k_{1 2} &=& \k_{0 3} \rightarrow \k_3 = 3 \nu_2 \sigma^4  
        + \left( -2 + 18 \nu_2 -{33\over 2} \nu_2^2
- 4\nu_3  \right. \nn \\ 
&+& \left. \nu_2^3 + 6 \nu_2 \nu_3 + {3\over 2} \nu_4 \right) \sigma^6 
        + \Or\left(\sigma^8\right) \nn 
\eea  
and thus,
\bea 
&& c_{1 2} \rightarrow S_3 ~=~ 3 \nu_2 +  
        \left(-2 + {15\over 2} \nu_2^2 -2 \nu_2^3 - 4 \nu_3 + {3\over 2} \nu_4 \right) \sigma^2  
%       + \Or\left(\sigma^4\right)
\nn \\ 
&& c_{1 3} = c_{2 2} \rightarrow S_4 ~=~ 12 \nu_2^2 + 4 \nu_3 \nn \\
&+&  \left(6 -36 \nu_2 + 15 \nu_2^2 + 60 \nu_2^3 - 15 \nu_2^4 - 4 \nu_3  
\right.  
\nn \\ 
&-& \left.  
20 \nu_2 \nu_3  - 6 \nu_2^2 \nu_3 - 5 \nu_4 + 18 \nu_2 \nu_4 + 2 \nu_5  
\right) \sigma^2  
%       + \Or\left(\sigma^4\right)
\nn ,
\eea 
which provides a 
reassuring check for our results for the 2-point moments.

\subsection{Smoothing Effects} 
 
Following FG98,
the smoothing effects in the moments of the density field in cells of 
given resolution $\lambda$,
can be easily incorporated.
In particular, smoothing simply changes the $\nu_n$ coefficients of the
SC model, Eq.(\ref{nusc}), 
by introducing {\em smoothing corrections} 
in terms of derivatives of the
logarithmic slope of the rms fluctuation, 
$\gamma = d \log\sigma^2/d \log\lambda$,   
\bea 
\nu_2 &\rightarrow& \nu_2 + {\gamma \over 3} \nn \\ 
\nu_3  &\rightarrow& \nu_3 - {\gamma \over 2} +
{3 \over 2} ~\nu_2~\gamma + {\gamma^2 \over 4} \nn \\
\nu_4 &\rightarrow& \nu_4 + {4 \over 3} ~\gamma - 4 ~\nu_2 ~\gamma
+ 2 ~\nu_2^2 ~\gamma + {8 \over 3} ~\nu_3 ~\gamma - {4 \over 3} ~\gamma^2
\nn \\
&+& {8 \over 3} ~\nu_2 ~\gamma^2 + {8 \over 27} ~\gamma^3 ,
\label{nusm} 
\eea   
where we have assumed for simplicity that 
higher-order derivatives of $\sigma$ are negligible
(see also Juszkiewicz et al. 1993). 
For an arbitrary power-law $P(k)$, the above results can be trivially 
generalized (see Appendix A of FG98).
Eqs.(\ref{c12scl})-(\ref{c22scl}) along with the 
expression for the smoothed vertices $\nu_n$, 
Eq.(\ref{nusm}), can be directly compared to Nbody measurements 
(see \S\ref{sec:nbody} below for a detailed discussion).
 
Alternatively, the PT solutions by B96 
for the  2-point cumulants at tree-level 
(neglecting again 
higher-order derivatives of $\sigma$) read:
\footnote{We have explicitly checked that the last term 
in Eq.(20) of B96 (neglected in the above expression) 
makes little difference for the large values of $r_{12}$ where this expression
applies.} 
\bea 
c_{1 2}^{PT} &=& {68 \over{21}} + {\gamma\over{3}} \nn \\ 
c_{1 3}^{PT} &=& {11710 \over{441}} + {61\over{7}}~\gamma   
+ {2\over{3}}~\gamma^2
\label{cijpt} 
\eea 

Substituting the smoothed vertices, Eq.(\ref{nusm}), into 
Eqs.(\ref{c12scl})-(\ref{c22scl}), 
one sees that both results differ in the smoothing effects.
For example, according to the SC model prediction, one finds
\bea 
c_{1 2} &=& {68 \over{21}} + {2\over{3}}~\gamma \nn \\ 
c_{1 3} &=& {11710\over 441} + {515\over42}~\gamma 
+ {17\over12}~\gamma^2
\label{c12sc0}
\eea 
which shows that the {\em 2-point skewness}, $c_{1 2}$, 
has an additional smoothing correction $\gamma/3$ with respect
to the PT results derived by B96, Eq.(\ref{cijpt}).
Similarly, the {\em 2-point kurtosis}, $c_{1 3}$, 
exhibits different smoothing effects in the SC 
model as compared to the PT results. 
Below we shall use Nbody simulations to 
investigate this discrepancy and discuss its interpretation.

\section{Comparison with Simulations} 
\label{sec:nbody}

In our numerical comparisons, we focus on
two classes of models
 that approximate well basic observations 
of galaxy clustering and, in particular, 
the APM galaxy Survey (Maddox \etal 1990). 

The first class of models are Cold Dark Matter (CDM) ones. In particular, 
we use the flat lambda CDM model ($\Lambda$CDM hereafter) with a
shape parameter $\Gamma=0.2$
and $\Omega_{m}=0.2$ ($\Omega_{\Lambda}=0.8$). We generally use
the outputs which have an amplitude of
mass fluctuations, $\sigma_8=1$, although
we have also considered other outputs in the range $\sigma_8=0.4-1.0$.  For
completeness we also consider the "old standard" CDM (SCDM from now on)
variant with $\Gamma=0.5$ and $\Omega_{m}=1$ ($\Omega_{\Lambda}=0$).  
Volume effects are assessed by comparing 
outputs from different box sizes in the range $300-600 \Mpc$.  

The second class of models considered have an APM-like linear power 
spectrum, so that, after evolution, 
it resembles closely the power spectrum
inferred from the APM galaxy catalogue
(see Baugh \& Gazta\~naga 1996). 
This model is normalized to $\sigma_8 \simeq
0.8$ corresponding to the mean redshift in the APM catalog ($z\simeq 0.15$).  

The small CDM
simulations are the ones in Baugh, Gazta\~naga \& Efstathiou (1995) and
the large ones (CDM and APM-like)
are from Gazta\~naga \& Baugh (1998), where more details can be
found. 
Errors are obtained from the dispersion in 10 (5) independent
realizations of the small (large) simulations. 
An additional advantage of using such simulations 
is that their higher-order moments
have been studied in detailed in the above cited references. 

Previous analyses of these simulations showed that 
the measured dark matter clustering, and in particular, its
1-point moments, are in good agreement with 
the {\em leading order} (also called tree-level) 
PT predictions,
$S_q \equiv \xibar_q/\xibar_2^{q-1}$. 
More recently, Fosalba \& Gazta\~naga (1998) found by using the SC model
that such agreement extends further into non-linear scales when
{\em next-to-leading order} PT predictions  
(the so-called loop corrections) are taken into account.

We have also run a new set of simulations with CDM shape $\Gamma=0.25$ but
with $\Omega_{m}=1$ (we call it $\Omega_m=1\times\Gamma=0.25$ 
CDM, where a CDM model
with these values would correspond to a model with a low Hubble constant $h
\simeq 0.25$. There is no need to make this correspondence,
however: it can be considered to be 
an $\Omega_m=1$ model with a $P(k)$ shape which is close to
that seen in observations).
These new simulations can be used to test the sensitivity to $\Omega_m$ with
independent of the shape of $P(k)$. They have a box size of
$300 \Mpc$ and twice the particle resolution to the previous sets ($N=200^3$
particles). Thus they can also be used to check for shot-noise and resolution
effects. The computation of gravitational dynamics was
 started at $z=20$ (instead of $z=10$
in the other CDM sets), so that they are less sensitive to possible 
Zeldovich Approximation
transients (see  Baugh, Gazta\~naga \& Efstathiou 1995,
Scoccimarro 1998).  There are 5 realizations 
of the $\sigma_8=0.5$ output.

\subsection{Estimators} 
\label{sec:estimator} 
 
The estimator of the density fluctuation at the $a$th pixel (or smoothed
point) is 
\beq 
\delta_a = {\rho_a \over {\langle \rho \rangle}}-1, 
\eeq 
where
$\rho_a$ is the mass density count in that pixel and $\langle \rho \rangle$ is
the overall mean value. The estimator for the 2-point moments can then be
generalized from Peebles \& Groth (1975) into: 
\beq 
{\hat m_{i,j}}(r) =
<\delta_a^i \delta_b^j > = {1\over N_r}\sum_{a,b} \delta_a^i \delta_b^j~
W_{ab}(r)~,
\label{1est} 
\eeq 
where $N_r = \sum_{a,b} W_{ab}(r)$ is the number of  
pairs of pixels (or smoothed fields) 
 at separation $r$ in the sample, and the  
window function  
$W_{ab}(r)=1$ if pixels $a$ and $b$ are separated by  
$|{\vec r}_a - {\vec r}_b| = r \pm dr$,   
and 0 otherwise.  
To obtain the reduced moments $k_{i,j}$ we have to use the
relations in equation (\ref{kijmij}).
In paper II of this series we  propose a variant
of this estimator that produces better results for smaller
samples. For the large periodic simulations we use here,
the above estimator is simpler to code and
works as well.

\subsection{The 2-point Function and the Linear Regime} 
\label{sec:xi2}
 
\begin{figure} %[t] 
\centering 
\centerline{\epsfysize=8.truecm %\epsfxsize=15.truecm  
\epsfbox{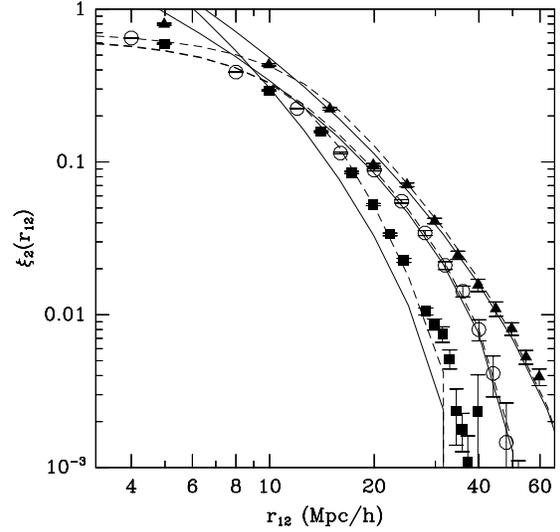}} 
%\figcaption{ 
\caption[junk]{The 2-point function $\xi_2 \equiv  
<\delta_1\delta_2>$  
as a function of the  
separation $r_{12}$ in Nbody simulations with a $\Gamma=0.2$ CDM  
(triangles),  $\Gamma=0.5$ CDM (squares) 
and APM-like (circles) spectrum. 
The continuous line shows the unsmoothed $\xi_2$  predicted 
by linear theory in each model.
The dashed line is the same linear prediction smoothed 
with the corresponding top-hat window. 
\label{x2r12}}
\end{figure} 
 
Figure \ref{x2r12} shows the comparison of the 2-point function
$\xi_2(r_{12})$ with linear theory.  The smoothing scale in this case are
$R_f=8 \Mpc$ (LCDM, SCDM) and $R_f=10\Mpc$
 (APM-like).  At large scales $r_{12}>
R_f$, the 2-point function agrees well with the (smoothed) linear theory
prediction.  We find that this is so for a wide ranges of smoothing scales,
even when the smoothing radius is smaller (and the corresponding variance
larger).  This shows that {\em the non-linear nature of small scale 
fluctuations does not affect significantly the large scale 
clustering, in agreement with PT predictions}.
 
On small scales, $r_{12}\simlt R_f$, 
the predictions follow closely 
the smoothed linear predictions (dashed line), but, as expected,  
there are non-linear corrections 
when the smoothing scale is such that the variance is 
larger than unity. 
 
Note that the 2-point function in the APM-like model drops 
more sharply than the low-density model (see Fig \ref{x2r12}). 
The SCDM  model drop is even steeper. At large scales, the shape is similar 
to the APM-like but it crosses zero at a smaller scale. 
At the scale where the two point function becomes zero the 
errors will be quite large, because of the relative effect 
of sampling fluctuations.  
This is also visible in the 2-point skewness, $c_{12}$, which is 
subject to large biases at this point (see below). 
 
These models are quite different in shape and we will 
further explore the dependence in $\gamma$ by considering 
different smoothing scales. To see how the properties vary for 
a fixed scale and shape ($\gamma$) we will also  
consider the statistics of 
different outputs (as parametrized by $\sigma_8$) 
from the same simulations. 
 
The agreement with linear PT shown in this section 
is a good test both for the code and estimator used. 
Higher-order 2-point moments are obtained from the same
codes as the 2-point function, the only difference being
the different powers of the density considered.

\begin{figure} %[t] 
\centering 
\centerline{\epsfysize=9truecm %\epsfxsize=15.truecm  
\epsfbox{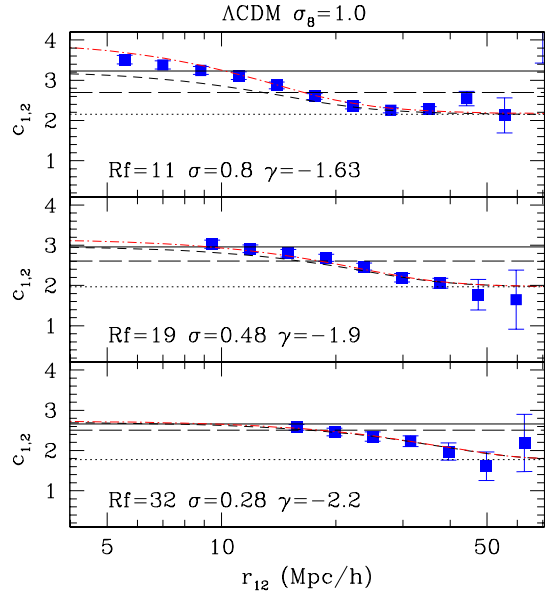}} 
\caption[junk]{The hierarchical coefficient $c_{1 2}$ in the  
$\Lambda$CDM model 
as a function of the  
separation $r_{12}$. Each panel shows results for
different smoothing scale $R_f$ as labeled in the Figure.
Symbols show  $c_{1 2}$ in the $\sigma_8=1.0$ 
output of the $\Lambda$CDM model,
with errors  corresponding to the variance in 5 realizations.  
The continuous line 
shows the leading order prediction for the skewness:  
$S_3 \equiv c_{1 2}(r_{12}=0)=34/7+\gamma$.  
The long-dashed line corresponds to the PT 
result  $c_{1 2} = 68/21+\gamma/3$ (B96). 
The short-dashed line corresponds to the SC 
prediction, Eq.(\ref{c12sc}), which tends to  
$c_{1 2} = 68/21+2\gamma/3$ (dotted-line) 
in the limit $\xi_2 \rightarrow 0$. The dotted-dashed 
lines correspond to the 1-loop SC prediction, Eq.(\ref{c12scl}). 
\label{c12c80dat6}} 
\end{figure}  

\begin{figure} %[t] 
\centering 
\centerline{\epsfysize=9truecm %\epsfxsize=15.truecm  
\epsfbox{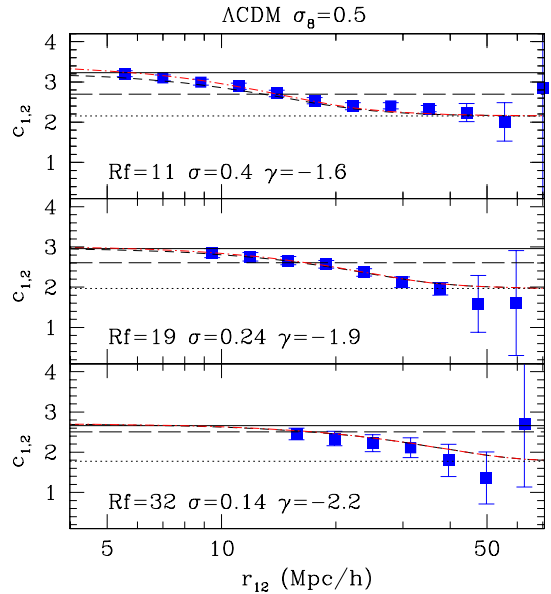}} 
\caption[junk]{Same as Fig.\ref{c12c80dat6} for the
$\sigma_8=0.5$ $\Lambda$CDM output.}
\label{c12c80dat3} 
\end{figure}  

\subsection{The 2-point Skewness in the Weakly Non-linear Regime}

Figures \ref{c12c80dat6}-\ref{c12c80dat3}  
show the 2-point skewness, $c_{1 2} \equiv  \k_{1 2} / (\xi_2
\sigma^2)$, in the 
weakly non-linear regime, that is when the smoothing 
scale is such that the variance is small: $\sigma \la 1$.  
The figures show a comparison with $\Lambda$CDM simulations for
a wide range of values of $\sigma =0.1-1.5$ ($\sigma^2 =0.02-2$)
corresponding to different 
scales, outputs and values of $\gamma$.
The long-dashed line corresponds to the tree-level
in PT prediction, Eq.(\ref{cijpt}). 
The short-dashed line corresponds to  
the leading order prediction
\footnote{Here by leading order in a PT calculation of the density
fluctuation field, $\delta$, we mean a term of order 
$\delta^0$, such as $\xi_2/\sigma^2$} 
for the 2-point skewness in the SC: 
\beq 
c_{1 2} =  2 ~\overline{\nu_2} + \overline{\nu_2} 
 ~{\xi_2\over{\sigma^2}} =
{68\over 21}+ {2\over3}~\gamma + \left({34\over21}+ {\gamma\over3}\right)
 ~{\xi_2\over{\sigma^2}}
 \label{c12sc} 
\eeq  
while the dotted lines gives the 
limit $\xi_2 \rightarrow 0$, e.g, Eq.(\ref{c12sc0}). 
The dot-dashed line includes the 1-loop SC correction 
(e.g, Eq.(\ref{c12scl})). 
Each panel is labeled with the smoothing scale $R_f$ and the  
corresponding linear rms $\sigma(R_f)$ and slope $\gamma$. 

In all cases we see there is a very good agreement with the leading order SC
prediction.  Note that in the limit $\xi_2 \rightarrow 0$ we recover the SC
value in Eq.(\ref{c12sc}) (dotted line) rather than the rigorous PT
prediction, Eq.(\ref{cijpt}) (long dashed line).  This is a somewhat
surprising result, as one would expect the rigorous PT prediction to be more
accurate than the SC approximation.  A possible explanation could be that the
regime of validity of PT is out of the dynamical range proved in our
simulations.  However this is unlikely to be the case as we have managed to
measure $c_{12}$ for scales as large as $R_f \ga 30 \Mpc$ and $r_{12} \ga 50
\Mpc$. On those scales the variance is $\sigma^2 < 0.02$ and $\xi_2$
approaches zero, where perturbative expansions of 2-point moments should yield
accurate predictions.

 On large scales the agreement with SC does not seem to depend on the
value of the variance at the smoothing scale, except when we approach $\sigma
\simeq 1$ and the perturbative approach itself breaks down 
(see \S\ref{sec:nonlinear}).

\begin{figure} %[t] 
\centering 
\centerline{\epsfysize=9truecm %\epsfxsize=15.truecm  
\epsfbox{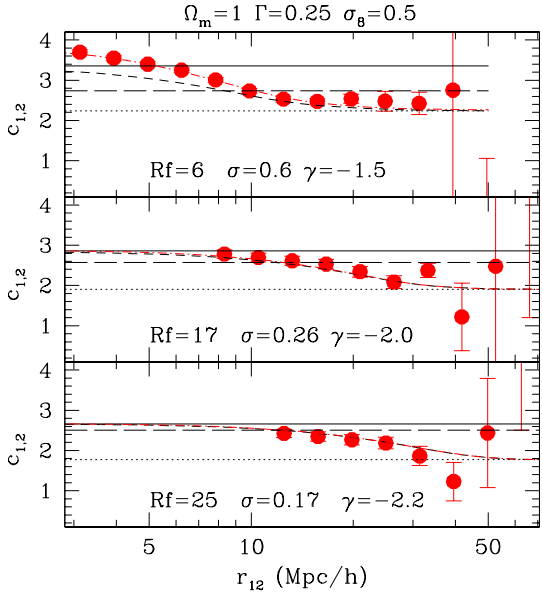}} 
\caption[junk]{Same as Fig.\ref{c12c80dat6} for
$\sigma_8=1.0$ $\Omega_m=1\times\Gamma=0.25$.}
\label{c12c75} 
\end{figure}  
 
Fig.\ref{c12c75} shows a comparison similar to  Fig.\ref{c12c80dat3} for the
$\Omega_m=1\times\Gamma=0.25$ model. The results are almost identical,
 indicating
that they are insensitive to the value of $\Omega_m$ and to the particle
resolution.

\begin{figure} %[t] 
\centering 
\centerline{\epsfysize=8.truecm %\epsfxsize=15.truecm  
\epsfbox{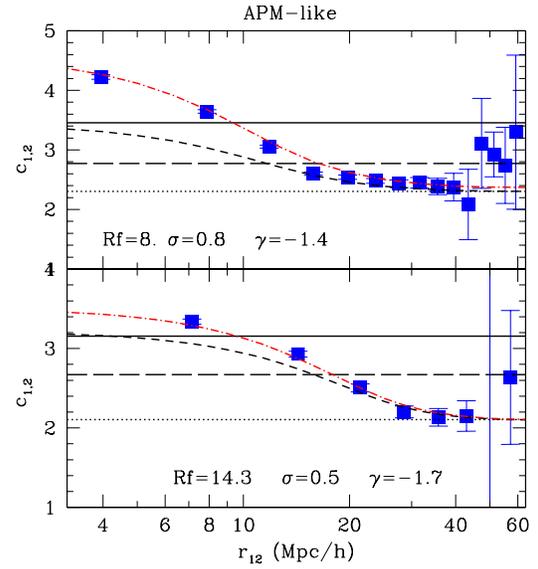}} 
\caption[junk]{Same as Figure \ref{c12c80dat6} for the 
APM-like model.}  
\label{c12c378} 
\end{figure}

\begin{figure} %[t] 
\centering 
\centerline{\epsfysize=8.truecm %\epsfxsize=15.truecm  
\epsfbox{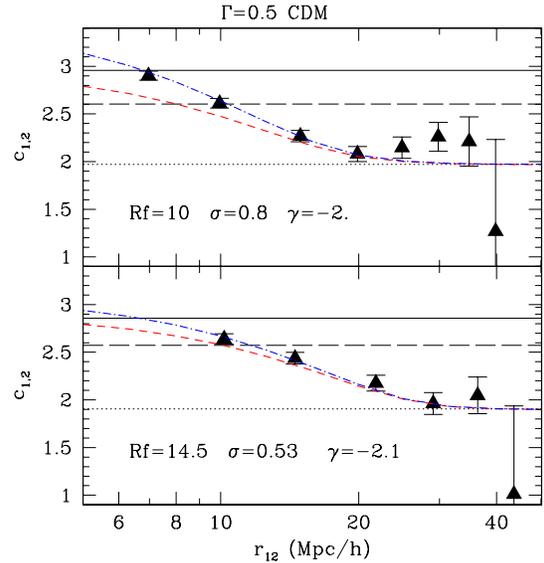}} 
\caption[junk]{Same as Figure \ref{c12c80dat6} for the 
$\Gamma=0.5$ CDM model.}  
\label{c12c189} 
\end{figure}

Fig. \ref{c12c378}, shows 
the corresponding comparison for the realistic APM-like model.  
It displays large fluctuations at  
$r_{12} \simeq 50 \Mpc$ 
which reflects the fact that the 2-point function in the APM-like 
model goes to zero on that scale. 
A similar problem arises around $r_{12} \simeq 40 \Mpc$ for the 
SCDM model in Fig \ref{c12c189}. As the two point function crosses 
zero quite sharply and the smoothing scale $R_f$ is large, the zero 
crossing affects a large range of scales  
 $r_{12} \simeq 40 \Mpc$ (typically a range $\pm  R_f$).

We can see Figure \ref{c12c378}  that the SC prediction gives 
a much better match to the simulations than the PT predictions, in 
agreement with the cases discussed previously. 
Both in Fig \ref{c12c378} and \ref{c12c189} we can see how the 
SC model including the next-to-leading order 
(i.e, 1-loop correction) accurately reproduces the higher-order
moments measured in simulations .

\subsection{The 2-point Skewness in the Non-linear Regime} 
\label{sec:nonlinear}
 
\begin{figure} %[t] 
\centering 
\centerline{\epsfysize=8.truecm %\epsfxsize=15.truecm  
\epsfbox{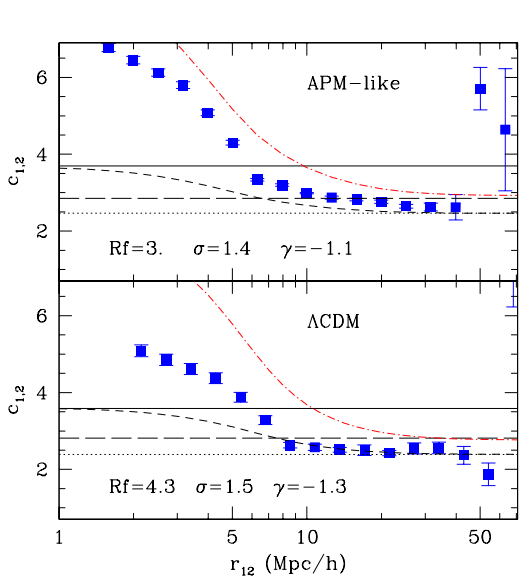}} 
%\figcaption{ 
\caption[junk]{Same as Figure \ref{c12c80dat6}, but for smaller smoothing 
scales (which means larger values of the variance). 
The top (bottom) panel shows the 
APM-like ($\Gamma=0.2$ CDM) model.}  
\label{c12sig2} 
\end{figure}

Figure \ref{c12sig2} shows a comparison for smaller smoothing radius. 
Here the variance at the smoothing scale $R_f \simeq 3-4$ is 
$\sigma^2 \simeq 2$.  
Notice that, as expected, significant 
departures from the tree-level prediction in
the SC model are observed on such small scales, i.e, 
when the variance largely exceeds unity. This is in agreement to what 
is already known for the skewness $S_3$. 
On the other hand, as can be seen in Fig \ref{c12sig2}, the SC predictions 
work well on large scales, where $\xi_2<1$ (e.g, for $r_{12}>8 \Mpc$), 
even when $\sigma^2>1$. 
This is similar to what we found for  $\xi_2$ in \S \ref{sec:xi2}, 
which follows well linear theory even when the smoothing scale is 
fully non-linear. 
 
For small smoothing radius $R_f$  
the errors become much smaller and the 
SC gives a perfect match to Nbody measurements 
at large $r_{12}$. This says that
the SC reproduces well the non-linear transition 
between small and 
large scales found in numerical simulations.

In general, for $\sigma \la 0.8$ the results agree quite well with the
perturbative SC model. In some cases, when $\gamma$ is large (flat slope) and
$\sigma \simeq 1$ the amplitudes $c_{12}$ tend to increase at all scales,
including at large $r_{12}$. This behavior is qualitatively predicted by
the perturbative SC expressions as given by Eq.(\ref{c12scl}), although 
in the non-linear regime, 
there is no reason to expect that
the term $\sim\sigma^2$ 
intrinsic to loop corrections in PT can accurately account for 
the gravitational evolution of the moments. 
We stress that the value of $\sigma$, where
deviations from SC prediction take place, depends on the value of the smoothing
parameter, $\gamma$, in agreement with the general trend found in FG98 for the
1-point moments: the larger
the smoothing correction (or the smaller $\gamma$), 
the smaller the non-linear contribution to the moments
(see Table 2 in FG98). This can also be seen at large scales ($r_{12} \ga 8
\Mpc$) by comparing the two panels in Fig.\ref{c12sig2}.

\subsection{The 2-point Kurtosis}

\begin{figure} %[t] 
\centering 
\centerline{\epsfysize=8.truecm %\epsfxsize=15.truecm  
\epsfbox{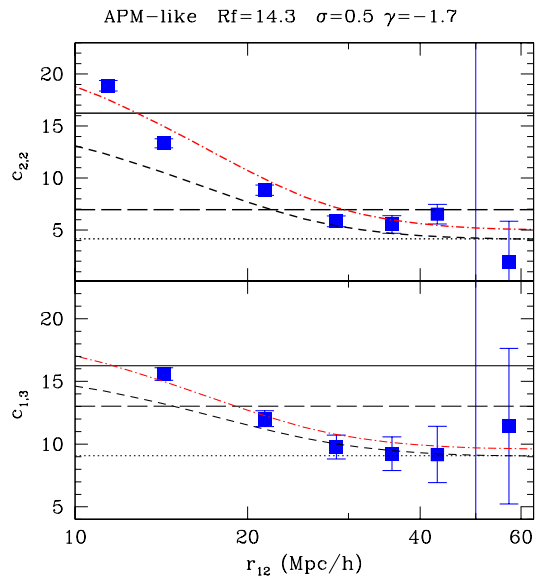}} 
%\figcaption{ 
\caption[junk]{The hierarchical coefficients $c_{2 2}$ (top panel) 
and $c_{1 3}$ (bottom panel)
in the APM-like model smoothed over a cell of $R_f=14.3$ ($\sigma=0.5$)
as a function of the cell separation $r_{12}$.  The continuous line 
shows the leading order prediction for the kurtosis:  
$S_4 \equiv c_{1 3}(r_{1 2}=0)  \equiv c_{2 2}(r_{1 2}=0)$.  
The long dashed line corresponds to the rigorous PT prediction.
The short-dashed line corresponds to the SC 
leading order predictions, which tend to the dotted-lines
in the limit $\xi_2 \rightarrow 0$. The dotted-dashed 
line corresponds to the 1-loop SC prediction as given by 
Eqs.(\ref{c13scl})-(\ref{c22scl}).
\label{c13c22rf14}} 
\end{figure}

\begin{figure} %[t] 
\centering 
\centerline{\epsfysize=8.truecm %\epsfxsize=15.truecm  
\epsfbox{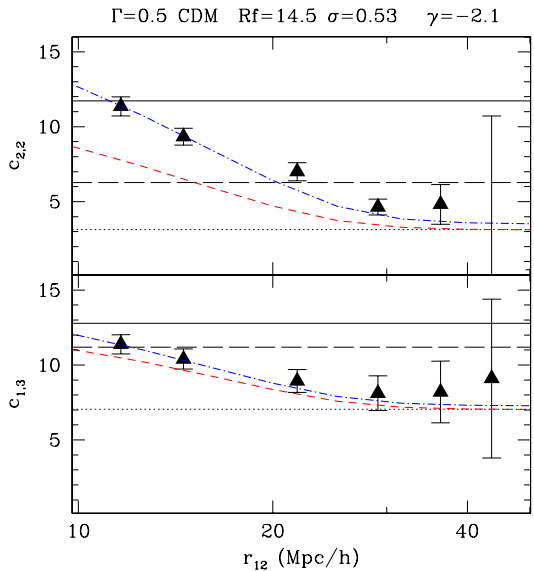}} 
\caption[junk]{Same as Fig.
\ref{c13c22rf14} for $\Gamma=0.5$ CDM.
\label{c13c22c189}}
\end{figure}

The analysis presented for the 2-point skewness above, can be 
straightforwardly extended to higher-order moments of the density field.
In particular, we shall compare the results in the SC model 
for the 2-point estimators of the volume averaged 4-point function, 
what we shall call 
the {\em 2-point kurtosis}, 
Eqs.(\ref{c13scl})-(\ref{c22scl}), with numerical simulations. 
This will allow us to
assess how general are the results found
from the analysis of the skewness.
Figure \ref{c13c22rf14} \& \ref{c13c22c189} compare
the SC prediction
for the normalized 2-point kurtosis, $c_{13}$ and
$c_{22}$, with the APM-like and CDM simulations, respectively.
Note how the limit for large scales (dotted lines)
is quite different for $c_{13} \rightarrow 6 \nu_2^2 + 3 \nu_3 $ 
than for $c_{22} \rightarrow  4 \nu_2^2 $. In both cases the SC
predictions are in agreement with simulations.

In general, for weakly non-linear scales,
$\sigma \la 0.5$, the results agree quite well with the
perturbative SC model. For larger values of $\sigma$ the amplitudes tend
to increase at all scales, including at large $r_{12}$. 

The observed
behavior is qualitatively predicted by the perturbative SC model,
Eqs.(\ref{c13scl})-(\ref{c22scl}),
although noticeable differences
appear when $\sigma \ga 0.5$.
This is due to the fact that 
the perturbative regime has a narrower domain of validity as one
considers higher-order moments of the density field.
This is so because loop corrections are 
relatively more important
for the 2-point kurtosis 
(see the difference between short-dashed and dot-dashed lines in 
Figs \ref{c13c22rf14} \& \ref{c13c22c189}) 
than for the 2-point skewness (see same lines in lower panels of 
Figs \ref{c12c378} \& \ref{c12c189}).

\section{Conclusions}
\label{sec:conc}

We have presented perturbative 
results for the 2-point moments of the dark matter
density field within the SC model.
In particular, we have derived expressions for the 2-point skewness and 
kurtosis in the weakly non-linear regime (see \S\ref{sec:ptresults}).

We have found  an excellent agreement between the SC model 
and CDM and APM-like simulations at all relevant scales. 
When the variance $\sigma^2 \la 0.5$ the
perturbative SC model is in full agreement (within the
error-bars) with
measurements of the 2-point skewness and kurtosis from numerical
simulations. 
When normalized to the second order statistics,
we have confirmed that our results are insensitive to the cosmological 
parameters or to the amplitude of the initial fluctuations.
We have tested the robustness of our findings 
against particle resolution and volume effects.

However, we have observed that the
domain of validity of the perturbative SC model gets narrower 
as one considers higher-order moments of the density field, as seem
in the analysis of the 2-point kurtosis (as compared to
that of the skewness).

For larger values of the variance of the matter density field, 
$\sigma^2$, the perturbative approach is observed to break down, as
expected. 
This is more extreme for smaller smoothing corrections. 
As we approach the strongly non-linear regime, $\sigma^2
\ga 2$, the leading order SC predictions 
match the simulations quite
well on scales $r_{12} > R_f$, while for $r_{12} \la R_f$ there are important
non-linear effects which cannot not accounted for 
within our perturbative approach, as one would expect.

Our numerical simulations fail to match the {\em leading order} analytic
predictions from perturbation theory (PT), derived by Bernardeau (1996, B96).
In the unsmoothed case, these calculations are identical to the SC results
presented here. But B96 found a different result for smoothed fields (ie
compare Eq.[\ref{cijpt}] with Eq.[\ref{c12sc0}]).  Our failure to reconcile
the simulations with the B96 predictions is puzzling.  We have managed to
extend our analysis to quite small values of the variance, i.e, $\sigma^2 \la
0.02$, and we have considered large simulation boxes (up to $600 \Mpc$) and
also different resolution cells. It is unlikely that larger (realistic)
simulations could resolve this problem as we are already probing scales where
the two point function goes to zero and starts oscilating.  On these larger
scales, both simulations and predictions become uncertain. The deviations we
find are quite significant, given the errors.  One possible future approach
could be to reduce the errors so as to be able to compare the predictions to
earlier simulation outputs. This would allow us to explore the regime where
$\sigma^2 \la 0.02$, while $\xi_2$ is still positive.

The idea that PT could fail for $\sigma^2 \simeq 0.02$ and start working for
$\sigma^2 \la 0.02$ is possible, but unlikely.  Experience with other PT
calculations and comparison with simulartions (e.g., with the bispectrum, see
Scocimarro etal 1998) indicates good convergence even for relatively large
variance $\sigma^2 \ga 0.1$.  It is reassuring that the perturbative SC
approach works so well for a wide range of (realistic) situations, 
such as those we have explored here, so that 
in practice we should use these predictions (rather
than B96) to compare to observations (see paper II).  Note that the proposal
by Szapudi (1998) to determine the bias with cumulant correlators will not
work in the regime we are exploring here (unlike B96 predictions, 
the 1-point and 2-point SC predictions have the 
same smoothing corrections).

The 2-point moments should provide a convenient tool to study the statistical
properties of gravitational clustering for fairly non-linear scales and
complicated survey geometries, as those probing the clustering of the
Ly$\alpha$ forest.  In this context, the perturbative SC predictions 
presented here provide a simple and novel way to test the gravitational
instability paradigm over a wide dynamical range.

\section*{Acknowledgments}

PF acknowledges support from an ESA fellowship and a CMBNET 
fellowship by the European Comission. E.G.
acknowledge support by grants from IEEC/CSIC and DGI/MCT(Spain) project
BFM2000-0810, and Acci\'on Especial ESP1998-1803-E.


\begin{thebibliography}{}

\def\refe {\par \hangindent=.7cm \hangafter=1 \noindent}
\def\apj {ApJ}
\def\na {Nature,}
\def\aap {A\&A}
\def\apjs{ApJS}
\def\mn {MNRAS}


\bibitem[Baugh, Gazta\~naga \& Efstathiou 1995]{bge95}
Baugh, C.M., \& Gazta\~naga, E., Efstathiou, G., 1995, \mn, 274, 1049

\bibitem[Baugh \& Gazta\~naga 1996]{bg96}
Baugh, C.M., \& Gazta\~naga, E., 1996, \mn, 280, 37

\bibitem[Bernardeau 1992]{b92} Bernardeau, F., 1992, \apj, 392, 1  

\bibitem[Bernardeau 1994]{b94} Bernardeau, F., 1994, \apj, 427, 51    

\bibitem[Bernardeau 1996]{b96} Bernardeau, F., 1996, \aap, 312, 11 (B96)

\bibitem[Davis \& Peebles, 1977]{davis77} Davis, M. \& Peebles,
P.J.E., 1977 \apjs, 34, 425

\bibitem[Fosalba \& Gazta\~naga 1998]{fg98}  Fosalba, P.,
Gazta\~naga, E., 1998, \mn,  301, 503 

%\bibitem[Fosalba \& Gazta\~naga 1998b]{fg98b}  Fosalba, P.,
%Gazta\~naga, E., 1998, \mn,  301, 534 

\bibitem[Fry 1984]{jf84} 
Fry, J. 1984, \apj, 277, L5

\bibitem[Gazta\~naga \& Croft 1999]{gc99} Gazta\~naga, E., Croft, R.A.C,
1999, \mn, 309, 885

\bibitem[Gazta\~naga \& Fosalba 1998]{gf98} Gazta\~naga, E., Fosalba,
P., 1998 \mn,  301, 524 

\bibitem[Gazta\~naga \& Yokoyama]{gy93} Gazta\~naga, E., Yokoyama, J.,
\apj, 1993, 403, 450

\bibitem[Gazta\~naga \& Scherrer 2001]{gs01} Gazta\~naga, E., Scherrer,
R.J, 2001, astro-ph/0105534, \mn, in press

\bibitem[Gazta\~naga et al.]{getal01} Gazta\~naga, E., \etal, 2001, in
preparation

%\bibitem[Gazta\~naga \& Frieman 1994]{gf94} Gazta\~naga, E.,
% Frieman, J.A., 1994, \apj, 437, L13 

\bibitem[Juszkiewicz et al. 1993]{rj93}
Juszkiewicz, R., Bouchet, F.R., \& Colombi, S., 1993, \apj, 412, L9  

\bibitem[Peebles 1980]{pee80} Peebles, P.J.E., 1980, The
Large--Scale Structure of the Universe,
Princeton University Press, Princeton (LSS)

\bibitem[Peebles 1993]{pee93} Peebles, P.J.E., 1993), Principles
of Physical Cosmology, Princeton University Press, Princeton

\bibitem[Scoccimarro \& Frieman 1996a]{sf96a} Scoccimarro, R.,
\& Frieman, J., 1996, \apjs., 105, 37

\bibitem[Scoccimarro \& Frieman 1996b]{sf96b} Scoccimarro, R.,
\& Frieman, J., 1996, \apj, 473, 620

\bibitem[Scoccimarro 1998]{sc98} Scoccimarro, R., 1998, \mn, 299, 1097


\bibitem[Scoccimarro 1998]{sc98} Scoccimarro, R.,
Colombi, S., Fry, J.~N., Frieman, J.~A., Hivon, E., \& Melott, A.\ 1998,
\apj, 496, 586

\bibitem[Szapudi(1998)]{1998MNRAS.300L..35S} Szapudi, I.\ 1998, \mn, 
300, L35 




\end{thebibliography}
\end{document}